\documentclass[a4paper,superscriptaddress,aps,prl,twocolumn,floatfix,citeautoscript,showpacs]{revtex4-1}
\usepackage{dcolumn}
\usepackage{amsmath}
\usepackage{graphicx}
\usepackage{latexsym}  
\usepackage{amsfonts}  
\usepackage{amssymb}
\usepackage{color}  
\usepackage{float} 

\DeclareGraphicsExtensions{.eps,.pdf,.png,.gif,.jpg}

\begin{document}  

\title{Unphysical and Physical Solutions in Many-Body Theories: from Weak to Strong Correlation}

\newcommand{\lsi}{Laboratoire des Solides Irradi\'es, \'Ecole Polytechnique, CNRS, CEA-DSM 
and European Theoretical Spectroscopy Facility (ETSF), 91128 Palaiseau, France}
\newcommand{\lpt}{Laboratoire de Physique Th\'eorique, CNRS, IRSAMC, Universit\'e Toulouse III - Paul Sabatier 
and European Theoretical Spectroscopy Facility (ETSF), 118 Route de Narbonne, 31062 Toulouse Cedex, France}
\newcommand{\lcpq}{Laboratoire de Chimie et Physique Quantiques, IRSAMC, Universit\'e Toulouse III - Paul Sabatier, CNRS 
and European Theoretical Spectroscopy Facility (ETSF), 118 Route de Narbonne, 31062 Toulouse Cedex, France}
\newcommand{\sor}{Sorbonne Universit\'es, UPMC Universit\'e Paris VI, UMR8112, LERMA, F-75005, Paris, France}
\newcommand{\lerma}{LERMA, Observatoire de Paris, PSL Research University, CNRS, UMR8112, F-75014, Paris, France}
\newcommand{\hum}{Humboldt-Universit\"at zu Berlin, Institut f\"ur Physik and IRIS Adlershof, 
and European Theoretical Spectroscopy Facility (ETSF), 12489 Berlin, Germany}

\affiliation{\sor}
\affiliation{\lerma}
\affiliation{\lsi}
\affiliation{\lpt}
\affiliation{\hum}
\affiliation{\lcpq}
\author{Adrian Stan}
\affiliation{\sor}
\affiliation{\lerma}
\affiliation{\lsi}
\author{Pina Romaniello}
\affiliation{\lpt}
\author{Santiago Rigamonti}
\affiliation{\hum}
\author{Lucia Reining}
\affiliation{\lsi}
\author{J. A. Berger}
\affiliation{\lcpq}

\date{\today}


%----------------------------------------------------------------------------------------------------
\begin{abstract}
Many-body theory is largely based on self-consistent equations that are constructed in terms of the physical quantity of
interest itself, for example the density. Therefore, the calculation of important properties such
as total energies or photoemission spectra requires the solution of non-linear equations that have unphysical and physical
solutions. In this work we show in which circumstances one runs into an
unphysical solution, and we indicate how one can overcome this problem. Moreover,
we solve the puzzle of when and why the interacting Green's function does
not unambiguously determine the underlying system, given in terms of its potential, or
non-interacting Green's function.
Our results are general since they originate from the fundamental structure of the equations. The
absorption spectrum of lithium fluoride is shown as one illustration, and
observations in the literature for some widely used models are explained by our
approach. Our findings apply to both the weak and strong-correlation regimes.
For the strong-correlation regime we show that one cannot use the
expressions that are obtained from standard perturbation theory, and we suggest
a different approach that is exact in the limit of strong interaction.
\end{abstract}
%----------------------------------------------------------------------------------------------------

\maketitle

In condensed-matter physics, important formalisms for predicting and understanding material properties, 
such as density-functional theory (DFT), many-body perturbation theory (MBPT), and dynamical mean-field theory (DMFT),
avoid the use of the full many-body wave function
and are instead based on simpler quantities, such as densities and Green's functions.
The one-body Green's function $G$, for example, yields expectation values of one-body operators, and the total 
energy. In particular, $G$ gives access to photoemission spectra, the most direct experimental observation of electronic structure.
$G$ is well defined as expectation value of particle addition and removal to the $N$-body ground state or 
thermal equilibrium. However, to use this definition one should know the many-body wave function, which is 
out of reach. A framework where the many-body wave function does not appear explicitly is provided by 
many-body perturbation theory \cite{fetterwal}, where the interacting Green's function $G$ is given as a functional of 
the non-interacting Green's function $G_0$ and the bare Coulomb interaction $v_c$. An important idea of MBPT 
is to avoid a possibly ill-behaved perturbation expansion of $G$ in terms of $v_c$ and $G_0$ using
Dyson equations. These are integral equations that describe the propagation of particles in terms of an effective
potential or interaction. For the one-body Green's function, for example, this effective potential, which is the kernel of the Dyson equation, 
is the self-energy $\Sigma$. The power of this approach resides in the fact that even a low-order approximation 
for $\Sigma$ yields contributions to all orders in $v_c$. Following Luttinger and Ward~\cite{LW},
$\Sigma$ is usually expressed as a functional of $G$ instead of $G_0$. However, this makes the Dyson equation 
non-linear, which leads to multiple solutions.~\cite{Lani,Tandetzky}

This is a very fundamental and general problem. It is different from usual
convergence problems, which are readily detected, for example from the oscillatory behavior of the
results; see the supplemental material for an example
of such a problem in the case of Hartree Fock (HF)~\cite{supmat}.
The appearance of fully converged, but unphysical results, instead, is
much more subtle and dangerous, and it has important consequences. 
It is the main topic of the present work.

We show that the presence of multiple solutions has an impact
that reaches far beyond numerical problems, and even points to cases
where the currently used strategies to derive approximations break down.
We develop our ideas using a model that represents the structure
of the exact theory. Calculations for a real system
and links to recent observations~\cite{Kozik_2014,Schaefer} demonstrate the potential
of this approach for analysis and prediction. In particular,
we answer several very general and important questions: 
{\textit{i}})
Is the problem of multiple solutions specific for certain cases, or is it 
a fundamental problem? 
Does it only appear for Green's functions or also in the framework of other well-established and widely used methods, 
like density-functional based approaches? 
{\textit{ii}}) Does it impact calculations for real
materials?; {\textit{iii}}) How can one detect and avoid unphysical
solutions?; {\textit{iv}}) Does the problem depend on the interaction strength, and
are there consequences for many-body
theory of strongly interacting materials?

% Starting from recent observations~\cite{Lani}, in this work we show that the presence of multiple solutions has consequences
% that reach far beyond numerical problems. We develop our ideas using a model that represents the structure of the exact 
% theory. Calculations for a real system and links to recent observations~\cite{Kozik_2014} demonstrate the potential of this model
% for analysis and prediction. In particular, we answer several questions of broad impact:
% \textit{{\textbf{i)}} How general is the problem of multiple solutions? Is it limited to self-energies
% or does it appear in other Dyson-like equations?; {\textbf{ii)}} Does it impact calculations for real materials?; 
% {\textbf{iii)}} How can one detect and avoid unphysical solutions?; {\textbf{iv)}} Are there any consequences for many-body 
% theory of strongly interacting materials?} 
% 
% 
% It is important to distinguish convergence problems from issues that have a deeper root in, and an impact on, the theoretical foundations.
% Convergence problems are readily detected from the oscillatory behavior of the results;
% see the supplemental material for an example of such a problem in the case of Hartree Fock (HF) \cite{supmat2}.
% A more fundamental problem, related to the structure 
% of the many-body equations, is the appearance of \emph{fully converged but unphysical} solutions.
% This is the main topic of the present work. 

To analyze the problem we use the so-called one-point model (OPM)~\cite{Molinari1,Molinari2,Pavlyukh}. 
This model is not system specific
and that can be solved exactly, such that the physical solution is well defined.
It represents important structural aspects of the many-body problem, while
collapsing all arguments of the Green's functions, self-energy, and the interaction to one point, 
making the equations scalar. In Ref.~\cite{Lani}, an approximate version of the OPM was used to discuss 
multiple solutions within the framework of the $GW$ approximation~\cite{PhysRev.139.A796} to the self-energy.

\begin{figure}[t]
\centering
\includegraphics[width=1.0\columnwidth]{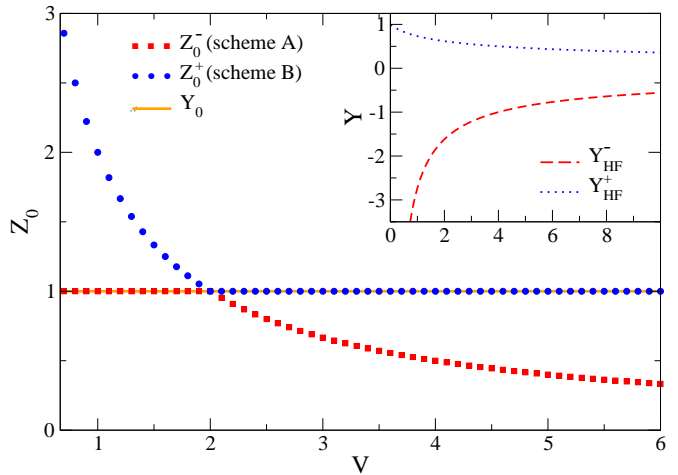}
\caption{(Color online) One-point model (OPM): $Z_0^{\pm}$ as a function of the interaction $V$.
Squares (red): $Z_0^-$ and solution of scheme (\textbf{A}); circles (blue): $Z_0^+$ and solution of scheme (\textbf{B});
continuous line (orange): the exact solution $Y_0$.
Inset: $Y_{\mathrm{HF}}^{\pm}$ as a function of the interaction $V$.}
\label{Fig:caos}
\end{figure}

In the present work we use the OPM without approximations, which simulates the full many-body problem.
The exact OPM Green's function was derived in Ref.~\cite{Berger_NJP14} from the one-point 
equivalent of the equation of motion of $G$, expressed as a functional  differential equation~\cite{martin59}.
The exact solution reads %The exact Green's function reads
\begin{equation}
 y[y_0,u] = \frac{y_0}{1+\frac12 uy_0^2}\,\,\,\,\,\,\,\textrm{and }\,\,\,\,\,\,\,\tilde s[y_0,u] = -\frac12 uy_0 ,
\label{Eqn:y_exact}
\end{equation}
where $y$, $y_0$, and $u$ represent $G$, $G_0$, and $v_c$, respectively.
The self-energy $\tilde s$  is determined from the Dyson equation $\tilde s[y_0,u]=y_0^{-1} - y^{-1}[y_0,u]$.
Here $\tilde s$ is given as a functional of the bare interaction $u$ and the noninteracting Green's function $y_0$. 
Usually, however, one works with the self-energy given as a functional of the dressed Green's function, $s[y,u]$.
Then the Dyson equation reads
\begin{equation}
y = y_0 + y_0 s[y,u] y .
\label{Eqn:Dyson_onepoint-y}
\end{equation}
This is, in general, a non-linear equation. 
We first consider the HF self-energy, which in the OPM is 
$s^{\mathrm{HF}}[y,u] = -\frac12 u y$.
Let us look at the map $G_0\to G$, {\it i.e.}, the usual case, where $y_0$ is set by the system, and one 
searches $y$. The Dyson equation has two solutions,
\begin{equation}
Y_{\mathrm{HF}}^{\pm} = \frac{1}{V}\left[ -1 \pm\sqrt{1+2V}\right],
\label{Eqn:G_HF}
\end{equation}
with the rescaled quantities $Y=y/y_0$ and $V=uy_0^2$. 
Here $Y_{\mathrm{HF}}^+$ is the physical solution, since it connects smoothly to $Y_0=1$ at $V=0$,
and $Y_{\mathrm{HF}}^-$ is an unphysical solution, that diverges for vanishing interaction.
Both are shown in the inset of Fig.~\ref{Fig:caos}.
In real problems Dyson equations are solved iteratively. Two possible iteration schemes are:
\begin{equation}
 Y^{(n+1)}= \frac{2}{2+VY^{(n)}} \,\,\mathrm{(\bf{I})} ; \,\,  Y^{(n+1)}= \frac{2}{VY^{(n)}}-\frac{2}{V} \,\, \mathrm{(\bf{II})}.
 \label{eq:iterate}
\end{equation}
While neither of the two schemes has convergence problems when iterating, only scheme (\textbf{I}) converges to 
the physical solution, whereas scheme (\textbf{II}) converges to the unphysical solution. 
This happens because the iteration leads to the continued fraction representation of the square root~\cite{Lani} in Eq.~(\ref{Eqn:G_HF}),
\begin{equation}
\sqrt{1+x}=1+\frac{x/2}{1+\frac{x/4}{1+\frac{x/4}{1+\cdots}}},
\end{equation}
for $x=2V$. The sign of the square root is determined by the continued fraction in the iterative procedure~\cite{mixing}.

The simple but general structure of the OPM suggests that the same picture should emerge for any Dyson-like equation. 
For example, optical properties and screening can be calculated by solving the Bethe-Salpeter equation (BSE) 
for the two-particle correlation function, using the $GW$ approximation for the self-energy~\cite{RevModPhys.74.601}.
The screened Coulomb interaction $W$ is calculated from the BSE, and is also part of its kernel.
Like in the above HF case, this makes the problem in principle self-consistent (see e.g. Ref.~\cite{PhysRevLett.85.2613}). %HF discussed above. 
Alternatively, one can use time-dependent density-functional theory (TDDFT), that obeys a similar Dyson-like equation for 
the reducible polarizability $\chi$~\cite{Petersilka}. 
Here we use TDDFT within the so-called bootstrap approximation proposed in Ref.~\cite{boot}. The corresponding Dyson equation for frequency $\omega$ reads
\begin{equation}
 \chi(\omega) = \chi_0(\omega) + \chi_0(\omega) \left [ v_c+\frac{1+v_c\chi(\omega=0)}{\chi_{0}(\omega=0)}\right ] \chi(\omega) ,
 \label{eq:tddft}
\end{equation}
where $\chi_0$ is the independent-particle polarizability.
We evaluate this for a real material with a long-range Coulomb interaction and \textit{ab initio} band structure.
Again, this equation has two solutions that can be obtained by two iteration schemes, analogous to those in Eq.~(\ref{eq:iterate})~\cite{supmat}.
\begin{figure}[t]
\centering
\includegraphics[angle=0, width=1.0\columnwidth]{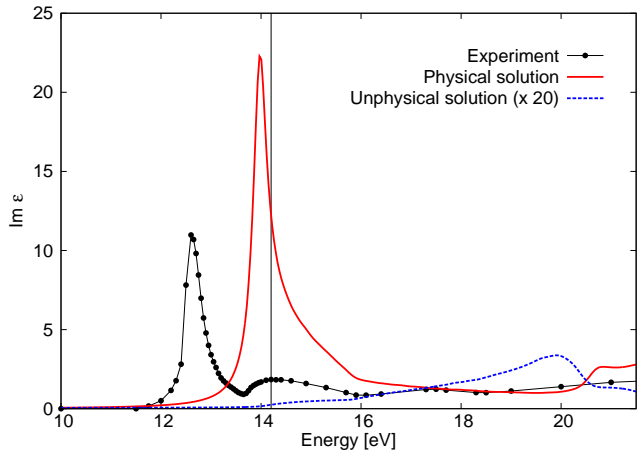}
\caption{(Color online) Optical absorption spectrum of LiF. 
Continuous line (red): physical solution; dashed line (blue): unphysical solution;
Dots (black): experiment~\cite{roessler}. 
The vertical line indicates the position of the quasiparticle gap.}
\label{fig:tddft}
\end{figure}
Figure~\ref{fig:tddft} shows the absorption spectrum of LiF obtained with the two iteration schemes, %of the TDDFT Dyson equation,
as well as the experimental result.
The experimental spectrum shows a strongly bound exciton with a binding energy of approximately 
1.4 eV ~\cite{roessler}.
The spectrum obtained from iterating Eq. (\ref{eq:tddft})
with the analogue of scheme (\textbf{I}) is qualitatively correct, since it also shows a strongly bound exciton.
The remaining discrepancies with respect to experiment are due to the approximate form of $f_{\rm xc}$ ~\cite{rigamonti_prl_2015}.
Instead, iterating Eq. (\ref{eq:tddft}) within scheme (\textbf{II})
makes the exciton disappear completely. This means that, in absence of experimental results, one risks to run into a wrong solution, which would 
make the theory non-predictive.
However, as we showed, the problem can be overcome using the appropriate iteration scheme (\textbf{I}). 
Note that optical properties can also be calculated from $\epsilon=1-v_c P$,
where the irreducible polarizability $P$ obeys a Dyson equation with kernel $f_{xc}=1/[(1-v_cP) \chi_0]$. In
this case, the appropriate scheme is scheme (\textbf{II}). This is explained in the supplemental material~\cite{supmat}.

So far we have looked at the map $G_0\rightarrow G$ (and $\chi_0\rightarrow \chi$). We now focus on the inverse map $G_0\leftarrow G$. 
This map is needed in problems of embedding, where one optimizes an auxiliary 
quantity $G_0$ in order to produce certain properties
of a real system (contained in $G$). The inverse map is also crucial when one wants to express a functional in terms of dressed instead of bare quantities.
The most prominent example is the Luttinger-Ward (LW) functional, where the self-energy is given 
in terms of $G$ instead of $G_0$ \cite{LW,Chitra,Potthoff,blochl_2013_density_matrix}.
For the LW functional to be properly defined, the map $G_0\leftarrow G$ should be unique.

Within the OPM, consider a system with the bare Green's function $y_0$, and with the exact, interacting 
Green's function $y$ given by Eq.~(\ref{Eqn:y_exact}).
We now fix $y$ and examine whether the inverse map $z_0 \leftarrow y$ unambiguously leads to $z_0=y_0$. 
With the exact self-energy $\tilde s[z_0] = -\frac12 u z_0$ of Eq.~(\ref{Eqn:y_exact}), the exact Dyson equation of 
this problem reads
\begin{equation}
z_0 = y + \frac{1}{2}u yz_0^2,
\label{Eqn:dyson_Y0}
\end{equation}
in which $y$ is known and $z_0$ is to be determined.   
This equation has again two solutions:
\begin{equation}
z_{0}^{\pm}=\frac{1}{uy}\left(1\pm\sqrt{1-2uy^2}\right) \Rightarrow Z_{0}^{\pm}=\frac{2+V\pm\sqrt{(2-V)^2}}{2V},
\label{Eqn:Z_0}
\end{equation}
where $Z_0 = z_0/y_0$ and we used Eq.\ ({\ref{Eqn:y_exact}}).  
The square root in Eq.~(\ref{Eqn:Z_0}) equals the absolute value $|2-V|$.
Because $2-V$ changes sign at $V=2$, the physical solution $Y_0=1$ is obtained by $Z_0^-$ for $V<2$ and by $Z_0^+$ for $V>2$ (see Fig.~\ref{Fig:caos}). 
In other words neither of the two solutions gives $Z_0=Y_0$ for all $V$. As a consequence \emph{one has to change sign in front of the square root} 
in Eq.~(\ref{Eqn:Z_0}) at $V=2$.  This has important consequences for the iterative solution of Eq.~(\ref{Eqn:dyson_Y0}):
because scheme (\textbf{I}) yields the square root with positive sign, to obtain the map $G_0 \leftarrow G$ we need \emph{two different iteration schemes}: 
one for $0<V<2$ and the other for $V>2$. This is different from the map $G_0 \rightarrow G$, where one solution gives the physical solution for all $V$, 
and hence a single iteration scheme suffices. 

The need to change iteration scheme is a serious problem. Indeed, Kozik \textit{et al.}~\cite{Kozik_2014} pointed out that different iteration schemes, 
applied to Hubbard 
and Anderson models, lead to different solutions which cross at a certain interaction.
Our OPM results provide the missing explanation:
keeping the labels (\textbf{A}) and (\textbf{B}) of \cite{Kozik_2014}, the two iteration schemes correspond to 
\begin{align}
\frac{1}{Z_0^{(n+1)}}=&1+\frac12 V (1-Z_0^{(n)})\quad\quad\quad\quad\quad \,\,\,\,  (\bf{A}),
\\
\frac{1}{Z_0^{(n+1)}}=&-1-\frac12 V (1-Z_0^{(n)}) + \frac{2}{Z_0^{(n)}} \quad  (\bf{B}).
\end{align}
We report the results in Fig.\ \ref{Fig:caos}.
Scheme (\textbf{A}) converges to the physical solution for $V<2$ and to the nonphysical solution for $V>2$.
Instead, scheme ($\textbf{B}$) converges to the nonphysical solution for $2/3<V<2$ and to the physical solution for $2<V<6$~\cite{nonconv}.
These results are strictly analogous to those obtained by Kozik \textit{et al.} for Hubbard and Anderson models. They can be understood from the fact that scheme (\textbf{A}) 
creates a continued fraction with positive square root, whereas 
in scheme (\textbf{B}) the sign of the continued fraction is changed.

This sign problem is \textit{a priori} a disaster because there is no unique prescription of how to avoid unphysical solutions.
The OPM highlights the reducible polarizability~\cite{Berger_NJP14}
\begin{equation}
\frac{\chi}{\chi_0}=2\frac{2-V}{(2+V)^2},
\end{equation}
as critical quantity that changes sign at the crossing $V=2$ \cite{polarizability}.
At the same time, for $V>2$ the perturbation expansion of $y$, in Eq.~(\ref{Eqn:y_exact}), diverges.
This is in line with Ref.~\cite{Schaefer} in which a breakdown of perturbation theory is linked
to an eigenvalue of the polarizability that crosses zero, becoming negative.
Our result confirms that one should inspect the reducible polarizability as a function of the interaction
to detect problems of perturbation theory.
%Whenever a qualitative change occurs, one should be mindful of the iteration scheme. 

Inverting a map between functionals requires a careful definition of their domain~\cite{Potthoff_EPJB,Eder_2014}.
The multiple solutions are the price to pay for the fact that we have not considered this definition in the above 
discussion.
This can be understood as follows: if there were two solutions for $G_0\leftarrow G$, one could obtain 
the same dressed Green's function from two different $G_0$ and, hence, from two different external potentials. 
Since the diagonal of $G$ is the density, the Hohenberg-Kohn theorem~\cite{HK} states that there can only be  
one external potential, and hence one $G_0$, corresponding to each $G$. This means that any additional 
solution $G_0$ is unphysical, in the sense that it cannot be constructed from the solution of a one-body 
Schr\"odinger equation.
Equivalently, it cannot be written as a sum of simple poles, each with a strength normalized to one. 
By imposing this condition, one can therefore eliminate unphysical solutions. 
In the OPM this trivially corresponds to the requirement $Z_0=1$, which already implies the solution.
In the supplemental material~\cite{supmat} we work out a more realistic case where the definition of the domain defines the solution. 
It should be noted that when $G_0$ is an embedding Green's function the discussion is more complicated, because 
one searches for a fictitious $G_0$ with a frequency dependence that can differ from that of a $G_0$ resulting from a 
static potential.
\begin{figure}[t]
\includegraphics[width=1.0\columnwidth]{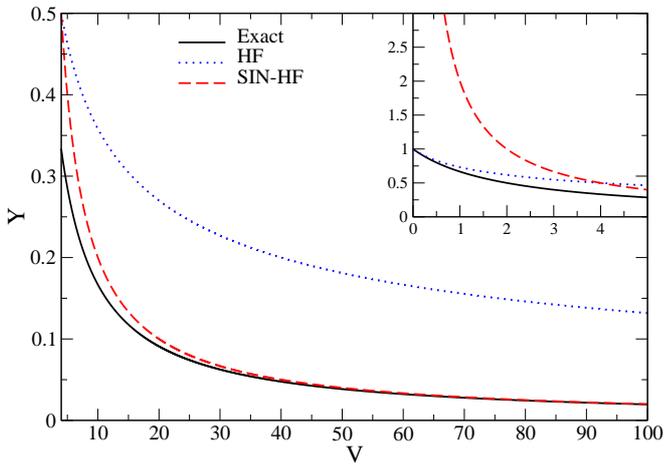}
\caption{(Color online) One-point model (OPM): $Y$ as a function of the interaction $V$ for $4<V<100$.
Solid line (black): exact solution; dotted line (blue): Hartree-Fock (HF); dashed line (red): strong-interaction HF (SIN-HF).
Inset: $Y$ as a function of $V$ for $0<V<5$.}
\label{Fig:SIN-HF}
\end{figure}

With the map $G_0 \leftarrow G$ one can construct the exact self-energy as a functional of $G$. 
Using Eq.\ (\ref{Eqn:Z_0}) in the exact self-energy given in Eq.~(\ref{Eqn:y_exact}) we obtain
\begin{align}
\label{Eqn:largevexp-1}
s^{\pm}[y,u] &= -\frac{1}{2y}\left(1\pm\sqrt{1-2uy^2}\right)
\\&=
-\frac{1}{2y}\mp \frac{1}{2y}\pm\frac{1}{2}u\left[y+\frac{uy^3}{2}+\frac{u^2y^5}{2}+...\right].
\label{Eqn:largevexp}
\end{align}
The Dyson equation with the two self-energies of Eq.~(\ref{Eqn:largevexp-1}) leads to two different Green's functions: 
the physical solution given in Eq.~(\ref{Eqn:y_exact}) is obtained from $s^-$ for $V<2$ and from $s^+$ for $V>2$, and $y=0$ from $s^+$ for any $V$.
Therefore, for weak interaction using the exact self-energy one obtains only one solution, the physical one, 
contrary to, \textit{e.g.}, the HF approximation.
We note that at the point where $s^+$ and $s^-$ meet (at $V=2$) the derivative $ds^{\pm}/dy$ diverges.
This could explain the divergence of $\delta\Sigma/\delta G$ observed in Ref.~\cite{Schaefer} for a 2D Hubbard model.
Note that this divergence occurs at the point where one of the eigenvalues of the polarizability crosses zero.

In Eq.~(\ref{Eqn:largevexp}) we Taylor expanded the square root.
The convergence radius is infinite, since $0\le 2uy^2\le 1$, as can be shown using Eq.~(\ref{Eqn:y_exact}).
Interestingly, the sum of the first two terms in (\ref{Eqn:largevexp}) (upper sign) is the first term of an expansion of the self-energy 
for strong interaction~\cite{Lani}. The remaining terms constitute an expansion in terms of a quantity that is proportional 
to $u$ and converges for all physical $y$. This means that 
one can use perturbation theory over the whole interaction range, but in two different ways for the two different regimes~\cite{largeint}.
To lowest order, this corresponds to 
HF, $s^{\mathrm{HF}} = -\frac{1}{2} uy$, for weak interaction,
and $s^{\mathrm{SIN-HF}} = -\frac{1}{y} - s^{\mathrm{HF}}$, for strong interaction.
We call this functional {\it strong-interaction HF} (SIN-HF). Both self-energies yield two solutions. 
We report the physical solution for these two approximations in Fig.~\ref{Fig:SIN-HF}.
While HF clearly fails for strong interaction, SIN-HF is exact in the strong interaction limit and performs well for $V>4$, while it is worse 
than HF for $V<4$. It is important to note that the physical SIN-HF solution is obtained for $V>1$ with the iteration scheme 
$1/Y^{(n+1)}=1/Y^{(n)} +\frac 12 VY^{(n)} -1$. 
Indeed, as it is also clear from the example of TDDFT, and discussed in the supplemental material~\cite{supmat}, 
the appropriate iteration scheme depends on the formulation of the problem. 
We suggest the OPM as a powerful tool
to examine which scheme one should use for a given problem and interaction range. 

In conclusion, we have demonstrated that with a simple but general one-point model
one can understand and solve structural problems of many-body perturbation theory.
In particular, one can use it sort out the multiple solutions of the non-linear 
Dyson equation by choosing the appropriate iteration scheme.
We have shown that for the map $G_0 \to G$ a single iteration scheme
suffices to obtain the physical solution for all interaction strenghts.
%the multiple solutions of the non-linear 
%Dyson equation can be sorted out by choosing the appropriate iteration scheme.
Instead, for the inverse map $G_0 \leftarrow G$ one has to change iteration scheme 
at the interaction strength at which the reducible polarizability changes sign and perturbation theory 
of $G$ in terms of $G_0$ starts to diverge.
Nonetheless, we have proved that even for strong interaction one can use a 
perturbative expression for the self-energy in terms of $G$, which differs 
from the usual LW functional. By presenting analogous results for 
real systems, and by comparing with numerical results in the literature, we have shown that these 
conclusions go far beyond the one-point model.
We expect that they will have an impact on many other questions in the domain of many-body physics.

\textit{Acknowledgments} We thank Antoine Georges for stimulating  discussions.
The research leading to these results has received funding from the
European Research Council under the European Union's Seventh Framework
Programme (FP/2007-2013) / ERC Grant Agreement n. 320971.
SR acknowledges support by the Einstein Foundation Berlin.

\bibliographystyle{apsrev}
\bibliography{manuscript}

\begin{thebibliography}{39}
\expandafter\ifx\csname natexlab\endcsname\relax\def\natexlab#1{#1}\fi
\expandafter\ifx\csname bibnamefont\endcsname\relax
  \def\bibnamefont#1{#1}\fi
\expandafter\ifx\csname bibfnamefont\endcsname\relax
  \def\bibfnamefont#1{#1}\fi
\expandafter\ifx\csname citenamefont\endcsname\relax
  \def\citenamefont#1{#1}\fi
\expandafter\ifx\csname url\endcsname\relax
  \def\url#1{\texttt{#1}}\fi
\expandafter\ifx\csname urlprefix\endcsname\relax\def\urlprefix{URL }\fi
\providecommand{\bibinfo}[2]{#2}
\providecommand{\eprint}[2][]{\url{#2}}

\bibitem[{\citenamefont{Fetter and Walecka}(2003)}]{fetterwal}
\bibinfo{author}{\bibfnamefont{A.}~\bibnamefont{Fetter}} \bibnamefont{and}
  \bibinfo{author}{\bibfnamefont{J.~D.} \bibnamefont{Walecka}},
  \emph{\bibinfo{title}{Quantum Theory of Many-Particle Systems}}
  (\bibinfo{publisher}{Dover publications}, \bibinfo{year}{2003}).

\bibitem[{\citenamefont{Luttinger and Ward}(1960)}]{LW}
\bibinfo{author}{\bibfnamefont{J.~M.} \bibnamefont{Luttinger}}
  \bibnamefont{and} \bibinfo{author}{\bibfnamefont{J.~C.} \bibnamefont{Ward}},
  \bibinfo{journal}{Phys. Rev.} \textbf{\bibinfo{volume}{118}},
  \bibinfo{pages}{1417} (\bibinfo{year}{1960}).

\bibitem[{\citenamefont{Lani et~al.}(2012)\citenamefont{Lani, Romaniello, and
  Reining}}]{Lani}
\bibinfo{author}{\bibfnamefont{G.}~\bibnamefont{Lani}},
  \bibinfo{author}{\bibfnamefont{P.}~\bibnamefont{Romaniello}},
  \bibnamefont{and} \bibinfo{author}{\bibfnamefont{L.}~\bibnamefont{Reining}},
  \bibinfo{journal}{New J. Phys.} \textbf{\bibinfo{volume}{14}},
  \bibinfo{pages}{013056} (\bibinfo{year}{2012}).

\bibitem[{\citenamefont{Tandetzky et~al.}(2012)\citenamefont{Tandetzky,
  Dewhurst, Sharma, and Gross}}]{Tandetzky}
\bibinfo{author}{\bibfnamefont{F.}~\bibnamefont{Tandetzky}},
  \bibinfo{author}{\bibfnamefont{J.~K.} \bibnamefont{Dewhurst}},
  \bibinfo{author}{\bibfnamefont{S.}~\bibnamefont{Sharma}}, \bibnamefont{and}
  \bibinfo{author}{\bibfnamefont{E.~K.~U.} \bibnamefont{Gross}},
  \bibinfo{journal}{arXiv:1205.4274.}  (\bibinfo{year}{2012}).

\bibitem[{sup()}]{supmat}
\bibinfo{note}{See Supplemental Material [URL inserted by publisher] which
  includes Refs.
  \cite{Cances_00,Hartree_57,Saunders_73,Pulay_80,Pulay_82,Cances_IJQC_00,Cances_02,exciting,perdewwang,exp_lif}}.

\bibitem[{\citenamefont{Kozik et~al.}(2015)\citenamefont{Kozik, Ferrero, and
  Georges}}]{Kozik_2014}
\bibinfo{author}{\bibfnamefont{E.}~\bibnamefont{Kozik}},
  \bibinfo{author}{\bibfnamefont{M.}~\bibnamefont{Ferrero}}, \bibnamefont{and}
  \bibinfo{author}{\bibfnamefont{A.}~\bibnamefont{Georges}},
  \bibinfo{journal}{Phys. Rev. Lett.} \textbf{\bibinfo{volume}{114}},
  \bibinfo{pages}{156402} (\bibinfo{year}{2015}).

\bibitem[{\citenamefont{Sch\"afer et~al.}(2013)\citenamefont{Sch\"afer,
  Rohringer, Gunnarsson, Ciuchi, Sangiovanni, and Toschi}}]{Schaefer}
\bibinfo{author}{\bibfnamefont{T.}~\bibnamefont{Sch\"afer}},
  \bibinfo{author}{\bibfnamefont{G.}~\bibnamefont{Rohringer}},
  \bibinfo{author}{\bibfnamefont{O.}~\bibnamefont{Gunnarsson}},
  \bibinfo{author}{\bibfnamefont{S.}~\bibnamefont{Ciuchi}},
  \bibinfo{author}{\bibfnamefont{G.}~\bibnamefont{Sangiovanni}},
  \bibnamefont{and} \bibinfo{author}{\bibfnamefont{A.}~\bibnamefont{Toschi}},
  \bibinfo{journal}{Phys. Rev. Lett.} \textbf{\bibinfo{volume}{110}},
  \bibinfo{pages}{246405} (\bibinfo{year}{2013}).

\bibitem[{\citenamefont{Molinari}(2005)}]{Molinari1}
\bibinfo{author}{\bibfnamefont{L.~G.} \bibnamefont{Molinari}},
  \bibinfo{journal}{Phys. Rev. B} \textbf{\bibinfo{volume}{71}},
  \bibinfo{pages}{113102} (\bibinfo{year}{2005}).

\bibitem[{\citenamefont{Molinari and Manini}(2006)}]{Molinari2}
\bibinfo{author}{\bibfnamefont{L.~G.} \bibnamefont{Molinari}} \bibnamefont{and}
  \bibinfo{author}{\bibfnamefont{N.}~\bibnamefont{Manini}},
  \bibinfo{journal}{Eur. Phys. J. B} \textbf{\bibinfo{volume}{51}},
  \bibinfo{pages}{331} (\bibinfo{year}{2006}).

\bibitem[{\citenamefont{Pavlyukh and H\"ubner}(2007)}]{Pavlyukh}
\bibinfo{author}{\bibfnamefont{Y.}~\bibnamefont{Pavlyukh}} \bibnamefont{and}
  \bibinfo{author}{\bibfnamefont{W.}~\bibnamefont{H\"ubner}},
  \bibinfo{journal}{J. Math. Phys.} \textbf{\bibinfo{volume}{48}},
  \bibinfo{eid}{052109} (\bibinfo{year}{2007}).

\bibitem[{\citenamefont{Hedin}(1965)}]{PhysRev.139.A796}
\bibinfo{author}{\bibfnamefont{L.}~\bibnamefont{Hedin}},
  \bibinfo{journal}{Phys. Rev.} \textbf{\bibinfo{volume}{139}},
  \bibinfo{pages}{A796} (\bibinfo{year}{1965}).

\bibitem[{\citenamefont{Berger et~al.}(2014)\citenamefont{Berger, Romaniello,
  Tandetzky, Mendoza, Brouder, and Reining}}]{Berger_NJP14}
\bibinfo{author}{\bibfnamefont{J.~A.} \bibnamefont{Berger}},
  \bibinfo{author}{\bibfnamefont{P.}~\bibnamefont{Romaniello}},
  \bibinfo{author}{\bibfnamefont{F.}~\bibnamefont{Tandetzky}},
  \bibinfo{author}{\bibfnamefont{B.~S.} \bibnamefont{Mendoza}},
  \bibinfo{author}{\bibfnamefont{C.}~\bibnamefont{Brouder}}, \bibnamefont{and}
  \bibinfo{author}{\bibfnamefont{L.}~\bibnamefont{Reining}},
  \bibinfo{journal}{New J. Phys.} \textbf{\bibinfo{volume}{16}},
  \bibinfo{pages}{113025} (\bibinfo{year}{2014}).

\bibitem[{\citenamefont{Martin and Schwinger}(1959)}]{martin59}
\bibinfo{author}{\bibfnamefont{P.~C.} \bibnamefont{Martin}} \bibnamefont{and}
  \bibinfo{author}{\bibfnamefont{J.}~\bibnamefont{Schwinger}},
  \bibinfo{journal}{Phys. Rev.} \textbf{\bibinfo{volume}{115}},
  \bibinfo{pages}{1342} (\bibinfo{year}{1959}).

\bibitem[{mix()}]{mixing}
\bibinfo{note}{We note that even using mixing, which is discussed in the
  Supplemental material, scheme (\textbf{II}) does not yield the physical
  solution for all $V$.}

\bibitem[{\citenamefont{Onida et~al.}(2002)\citenamefont{Onida, Reining, and
  Rubio}}]{RevModPhys.74.601}
\bibinfo{author}{\bibfnamefont{G.}~\bibnamefont{Onida}},
  \bibinfo{author}{\bibfnamefont{L.}~\bibnamefont{Reining}}, \bibnamefont{and}
  \bibinfo{author}{\bibfnamefont{A.}~\bibnamefont{Rubio}},
  \bibinfo{journal}{Rev. Mod. Phys.} \textbf{\bibinfo{volume}{74}},
  \bibinfo{pages}{601} (\bibinfo{year}{2002}).

\bibitem[{\citenamefont{Chang et~al.}(2000)\citenamefont{Chang, Rohlfing, and
  Louie}}]{PhysRevLett.85.2613}
\bibinfo{author}{\bibfnamefont{E.~K.} \bibnamefont{Chang}},
  \bibinfo{author}{\bibfnamefont{M.}~\bibnamefont{Rohlfing}}, \bibnamefont{and}
  \bibinfo{author}{\bibfnamefont{S.~G.} \bibnamefont{Louie}},
  \bibinfo{journal}{Phys. Rev. Lett.} \textbf{\bibinfo{volume}{85}},
  \bibinfo{pages}{2613} (\bibinfo{year}{2000}).

\bibitem[{\citenamefont{Gross et~al.}(1996)\citenamefont{Gross, Dobson, and
  Petersilka}}]{Petersilka}
\bibinfo{author}{\bibfnamefont{E.~K.~U.} \bibnamefont{Gross}},
  \bibinfo{author}{\bibfnamefont{J.~F.} \bibnamefont{Dobson}},
  \bibnamefont{and}
  \bibinfo{author}{\bibfnamefont{M.}~\bibnamefont{Petersilka}},
  \bibinfo{journal}{Top. Curr. Chem.} \textbf{\bibinfo{volume}{181}},
  \bibinfo{pages}{81} (\bibinfo{year}{1996}).

\bibitem[{\citenamefont{Sharma et~al.}(2011)\citenamefont{Sharma, Dewhurst,
  Sanna, and Gross}}]{boot}
\bibinfo{author}{\bibfnamefont{S.}~\bibnamefont{Sharma}},
  \bibinfo{author}{\bibfnamefont{J.~K.} \bibnamefont{Dewhurst}},
  \bibinfo{author}{\bibfnamefont{A.}~\bibnamefont{Sanna}}, \bibnamefont{and}
  \bibinfo{author}{\bibfnamefont{E.~K.~U.} \bibnamefont{Gross}},
  \bibinfo{journal}{Phys. Rev. Lett.} \textbf{\bibinfo{volume}{107}},
  \bibinfo{pages}{186401} (\bibinfo{year}{2011}).

\bibitem[{\citenamefont{Roessler and Walker}(1967)}]{roessler}
\bibinfo{author}{\bibfnamefont{D.~M.} \bibnamefont{Roessler}} \bibnamefont{and}
  \bibinfo{author}{\bibfnamefont{W.~C.} \bibnamefont{Walker}},
  \bibinfo{journal}{J. Opt. Soc. Am.} \textbf{\bibinfo{volume}{57}},
  \bibinfo{pages}{835} (\bibinfo{year}{1967}).

\bibitem[{\citenamefont{Rigamonti et~al.}(2015)\citenamefont{Rigamonti, Botti,
  Veniard, Draxl, Reining, and Sottile}}]{rigamonti_prl_2015}
\bibinfo{author}{\bibfnamefont{S.}~\bibnamefont{Rigamonti}},
  \bibinfo{author}{\bibfnamefont{S.}~\bibnamefont{Botti}},
  \bibinfo{author}{\bibfnamefont{V.}~\bibnamefont{Veniard}},
  \bibinfo{author}{\bibfnamefont{C.}~\bibnamefont{Draxl}},
  \bibinfo{author}{\bibfnamefont{L.}~\bibnamefont{Reining}}, \bibnamefont{and}
  \bibinfo{author}{\bibfnamefont{F.}~\bibnamefont{Sottile}},
  \bibinfo{journal}{Phys. Rev. Lett.} \textbf{\bibinfo{volume}{114}},
  \bibinfo{pages}{146402} (\bibinfo{year}{2015}).

\bibitem[{\citenamefont{Chitra and Kotliar}(2001)}]{Chitra}
\bibinfo{author}{\bibfnamefont{R.}~\bibnamefont{Chitra}} \bibnamefont{and}
  \bibinfo{author}{\bibfnamefont{G.}~\bibnamefont{Kotliar}},
  \bibinfo{journal}{Phys. Rev. B} \textbf{\bibinfo{volume}{63}},
  \bibinfo{pages}{115110} (\bibinfo{year}{2001}).

\bibitem[{\citenamefont{Potthoff}(2006)}]{Potthoff}
\bibinfo{author}{\bibfnamefont{M.}~\bibnamefont{Potthoff}},
  \bibinfo{journal}{Condensed Matter Physics} \textbf{\bibinfo{volume}{9}},
  \bibinfo{pages}{557} (\bibinfo{year}{2006}).

\bibitem[{\citenamefont{Bl\"ochl et~al.}(2013)\citenamefont{Bl\"ochl, Pruschke,
  and Potthoff}}]{blochl_2013_density_matrix}
\bibinfo{author}{\bibfnamefont{P.~E.} \bibnamefont{Bl\"ochl}},
  \bibinfo{author}{\bibfnamefont{T.}~\bibnamefont{Pruschke}}, \bibnamefont{and}
  \bibinfo{author}{\bibfnamefont{M.}~\bibnamefont{Potthoff}},
  \bibinfo{journal}{Phys. Rev. B} \textbf{\bibinfo{volume}{88}},
  \bibinfo{pages}{205139} (\bibinfo{year}{2013}).

\bibitem[{non()}]{nonconv}
\bibinfo{note}{For $V<2/3$ and $V>6$ scheme (\textbf{B}) does not converge.}

\bibitem[{pol()}]{polarizability}
\bibinfo{note}{The polarizability is defined as the derivative of $G$ with
  respect to an external potential. Therefore there is only one branch.}

\bibitem[{\citenamefont{Potthoff}(2003)}]{Potthoff_EPJB}
\bibinfo{author}{\bibfnamefont{M.}~\bibnamefont{Potthoff}},
  \bibinfo{journal}{Eur. Phys. J. B} \textbf{\bibinfo{volume}{32}},
  \bibinfo{pages}{429} (\bibinfo{year}{2003}).

\bibitem[{\citenamefont{Eder}(2014)}]{Eder_2014}
\bibinfo{author}{\bibfnamefont{R.}~\bibnamefont{Eder}},
  \bibinfo{journal}{arXiv:1407.6599}  (\bibinfo{year}{2014}).

\bibitem[{\citenamefont{Hohenberg and Kohn}(1964)}]{HK}
\bibinfo{author}{\bibfnamefont{P.}~\bibnamefont{Hohenberg}} \bibnamefont{and}
  \bibinfo{author}{\bibfnamefont{W.}~\bibnamefont{Kohn}},
  \bibinfo{journal}{Phys. Rev.} \textbf{\bibinfo{volume}{136}},
  \bibinfo{pages}{B864} (\bibinfo{year}{1964}).

\bibitem[{lar()}]{largeint}
\bibinfo{note}{Note that since $y\rightarrow 0$ for $u\rightarrow\infty$ the
  convergence of perturbation theory improves with increasing $u$.}

\bibitem[{\citenamefont{{Canc{\`e}s, Eric} and {Le Bris,
  Claude}}(2000)}]{Cances_00}
\bibinfo{author}{\bibnamefont{{Canc{\`e}s, Eric}}} \bibnamefont{and}
  \bibinfo{author}{\bibnamefont{{Le Bris, Claude}}}, \bibinfo{journal}{ESAIM:
  M2AN} \textbf{\bibinfo{volume}{34}}, \bibinfo{pages}{749}
  (\bibinfo{year}{2000}).

\bibitem[{\citenamefont{Hartree}(1957)}]{Hartree_57}
\bibinfo{author}{\bibfnamefont{D.~R.} \bibnamefont{Hartree}},
  \emph{\bibinfo{title}{The calculation of atomic structures}}
  (\bibinfo{publisher}{John Wiley and Sons, Inc.}, \bibinfo{address}{New York},
  \bibinfo{year}{1957}).

\bibitem[{\citenamefont{Saunders and Hillier}(1973)}]{Saunders_73}
\bibinfo{author}{\bibfnamefont{V.~R.} \bibnamefont{Saunders}} \bibnamefont{and}
  \bibinfo{author}{\bibfnamefont{I.~H.} \bibnamefont{Hillier}},
  \bibinfo{journal}{Int. J. Quantum Chem.} \textbf{\bibinfo{volume}{7}},
  \bibinfo{pages}{699} (\bibinfo{year}{1973}).

\bibitem[{\citenamefont{Pulay}(1980)}]{Pulay_80}
\bibinfo{author}{\bibfnamefont{P.}~\bibnamefont{Pulay}},
  \bibinfo{journal}{Chem. Phys. Lett.} \textbf{\bibinfo{volume}{73}},
  \bibinfo{pages}{393 } (\bibinfo{year}{1980}).

\bibitem[{\citenamefont{Pulay}(1982)}]{Pulay_82}
\bibinfo{author}{\bibfnamefont{P.}~\bibnamefont{Pulay}}, \bibinfo{journal}{J.
  Comp. Chem.} \textbf{\bibinfo{volume}{3}}, \bibinfo{pages}{556}
  (\bibinfo{year}{1982}).

\bibitem[{\citenamefont{Canc{\`e}s and Le~Bris}(2000)}]{Cances_IJQC_00}
\bibinfo{author}{\bibfnamefont{E.}~\bibnamefont{Canc{\`e}s}} \bibnamefont{and}
  \bibinfo{author}{\bibfnamefont{C.}~\bibnamefont{Le~Bris}},
  \bibinfo{journal}{Int. J. Quantum Chem.} \textbf{\bibinfo{volume}{79}},
  \bibinfo{pages}{82} (\bibinfo{year}{2000}).

\bibitem[{\citenamefont{Kudin et~al.}(2002)\citenamefont{Kudin, Scuseria, and
  Canc{\`e}s}}]{Cances_02}
\bibinfo{author}{\bibfnamefont{K.~N.} \bibnamefont{Kudin}},
  \bibinfo{author}{\bibfnamefont{G.~E.} \bibnamefont{Scuseria}},
  \bibnamefont{and}
  \bibinfo{author}{\bibfnamefont{E.}~\bibnamefont{Canc{\`e}s}},
  \bibinfo{journal}{J. Chem. Phys.} \textbf{\bibinfo{volume}{116}},
  \bibinfo{pages}{8255} (\bibinfo{year}{2002}).

\bibitem[{\citenamefont{Gulans et~al.}(2014)\citenamefont{Gulans, Kontur,
  Meisenbichler, Nabok, Pavone, Rigamonti, Sagmeister, Werner, and
  Draxl}}]{exciting}
\bibinfo{author}{\bibfnamefont{A.}~\bibnamefont{Gulans}},
  \bibinfo{author}{\bibfnamefont{S.}~\bibnamefont{Kontur}},
  \bibinfo{author}{\bibfnamefont{C.}~\bibnamefont{Meisenbichler}},
  \bibinfo{author}{\bibfnamefont{D.}~\bibnamefont{Nabok}},
  \bibinfo{author}{\bibfnamefont{P.}~\bibnamefont{Pavone}},
  \bibinfo{author}{\bibfnamefont{S.}~\bibnamefont{Rigamonti}},
  \bibinfo{author}{\bibfnamefont{S.}~\bibnamefont{Sagmeister}},
  \bibinfo{author}{\bibfnamefont{U.}~\bibnamefont{Werner}}, \bibnamefont{and}
  \bibinfo{author}{\bibfnamefont{C.}~\bibnamefont{Draxl}},
  \bibinfo{journal}{Journal of Physics: Condensed Matter}
  \textbf{\bibinfo{volume}{26}}, \bibinfo{pages}{363202}
  (\bibinfo{year}{2014}).

\bibitem[{\citenamefont{Perdew and Wang}(1986)}]{perdewwang}
\bibinfo{author}{\bibfnamefont{J.~P.} \bibnamefont{Perdew}} \bibnamefont{and}
  \bibinfo{author}{\bibfnamefont{Y.}~\bibnamefont{Wang}},
  \bibinfo{journal}{Phys. Rev. B} \textbf{\bibinfo{volume}{33}},
  \bibinfo{pages}{8800} (\bibinfo{year}{1986}).

\bibitem[{\citenamefont{Piacentini et~al.}(1976)\citenamefont{Piacentini,
  Lynch, and Olson}}]{exp_lif}
\bibinfo{author}{\bibfnamefont{M.}~\bibnamefont{Piacentini}},
  \bibinfo{author}{\bibfnamefont{D.~W.} \bibnamefont{Lynch}}, \bibnamefont{and}
  \bibinfo{author}{\bibfnamefont{C.~G.} \bibnamefont{Olson}},
  \bibinfo{journal}{Phys. Rev. B} \textbf{\bibinfo{volume}{13}},
  \bibinfo{pages}{5530} (\bibinfo{year}{1976}).

\end{thebibliography}

\end{document}